\global\def\draftcontrol{0}

   \def\versionno{ bulk bound violation }

\catcode`\@=11

\expandafter\ifx\csname draftcontrol\endcsname\relax\global\def\draftcontrol{0}
\fi

{\count255=\time\divide\count255 by 60
\xdef\hourmin{\number\count255}
\multiply\count255 by-60\advance\count255 by\time
\xdef\hourmin{\hourmin:\ifnum\count255<10 0\fi\the\count255}}
\def\draftdate{\number\month/\number\day/\number\year\ \ \ \hourmin }

\newcommand\makepapertitle{\par
  \begingroup
    \renewcommand\thefootnote{\@fnsymbol\c@footnote}%
    \def\@makefnmark{\rlap{\@textsuperscript{\normalfont\@thefnmark}}}%
    \long\def\@makefntext##1{\parindent 1em\noindent
            \hb@xt@1.8em{%
                \hss\@textsuperscript{\normalfont\@thefnmark}}##1}%
     \newpage
     \global\@topnum\z@   
     \@makepapertitle
     \thispagestyle{empty}\@thanks
  \endgroup
  \setcounter{footnote}{0}%
  \global\let\thanks\relax
  \global\let\makepapertitle\relax
  \global\let\@makepapertitle\relax
  \global\let\@thanks\@empty
  \global\let\@author\@empty
  \global\let\@date\@empty
  \global\let\@title\@empty
  \global\let\title\relax
  \global\let\author\relax
  \global\let\date\relax
  \global\let\and\relax
  \def\version{\let\version\@version\@gobble}
}
\def\@makepapertitle{%
  \newpage
   \ifnum\draftcontrol=1 {}
   \version\versionno
   \vskip 3em%
   \else
   \hfill\hbox to 3cm {\parbox{4cm}{\@pubnum}\hss}%
   \vskip 3em%
   \fi
   \begin{center}%
   \let \footnote \thanks
     {\LARGE {\@title}}%
     \vskip 1.5em%
     {\normalsize
       \lineskip .5em%
       \begin{tabular}[t]{c}%
         \@author
       \end{tabular}\par}%
     \vskip 1.5em%
     {\@bstract}%
     \end{center}%
     \vskip 1.5em
     \@date%
   \par
}

\gdef\@pubnum{}
\def\pubnum#1{%
  \gdef\@pubnum{#1}}

\gdef\@bstract{}
\def\Abstract#1{%
  \gdef\@bstract{%
   \parbox{\textwidth-0pc}{%
   \centerline{\bf Abstract}\penalty1000%
\kern.2cm%
\noindent
\renewcommand\baselinestretch{1.0}%
{#1}}}
}

\def\ps@paper{\let\@mkboth\@gobbletwo%
     \ifnum\draftcontrol=1
    \def\@oddfoot{\hbox to \textwidth{\tiny \versionno \hfil\tiny\draftdate}%
    \hskip -\textwidth \hbox to \textwidth{\hfil\rm\thepage\hfil}}%
     \else\def\@oddfoot{\hbox to \textwidth{\hfil\rm\thepage\hfil}}
     \fi
     \let\@evenfoot\@oddfoot
}

\def\body{\clearpage
          \pagestyle{paper}
    }

\def\@version#1{\ifnum\draftcontrol=1
\typeout{}\typeout{#1}\typeout{}
\vskip3mm\centerline{\hbox{\fbox{\normalsize{\tt DRAFT -- #1 -- }
                   {\draftdate}}}}\vskip3mm
\fi}
\let\version\@version
\long\def\eqlabel#1{\ifnum\draftcontrol=1
                    \tag@false  
                    \tag*{(\theequation) \hbox to -0.2cm{\hspace{0cm}\small{#1}\hss}}
                    \refstepcounter{equation}
                    \edef\@currentlabel{\theequation}
                    \ltx@label{#1}          
                    \else
                    \label{#1}
                    \fi
                    }
\let\st@bibitem\@bibitem
\let\st@lbibitem\@lbibitem
\ifnum\draftcontrol=1
  \def\@bibitem#1{%
    \st@bibitem{#1}\a@@label{#1}\ignorespaces}
  \def\@lbibitem[#1]#2{%
    \st@lbibitem[#1]{#2}\a@@label{#2}\ignorespaces}
  \def\a@@label#1{%
    \gdef\a@lab{\smash{\normalfont\small#1}}
    \ifvmode
      \if@inlabel
        \global\setbox\@labels\hbox{%
          \llap{\a@lab\let\a@lab\relax
                \kern\@totalleftmargin\kern\marginparsep}%
          \box\@labels}%
      \fi
    \fi}
\fi

\documentclass[12pt,letterpaper]{article}

\usepackage{amsmath,amssymb,array,calc,epsfig,rotating,bm}
\usepackage[sort]{cite}
\usepackage{graphicx}
\usepackage{psfrag,verbatim}

\ifnum\draftcontrol=1
\fi

\renewcommand\baselinestretch{1.25}
\renewcommand\section{\@startsection {section}{1}{\z@}%
                                   {-3.5ex \@plus -1ex \@minus -.2ex}%
                                   {2.3ex \@plus.2ex}%
                                   {\normalfont\large\bfseries}}
\renewcommand\subsection{\@startsection{subsection}{2}{\z@}%
                                   {-3.25ex\@plus -1ex \@minus -.2ex}%
                                   {1.5ex \@plus .2ex}%
                                   {\normalfont\normalsize\bfseries}}
\renewcommand\subsubsection{\@startsection{subsubsection}{3}{\z@}%
                                   {-3.25ex\@plus -1ex \@minus -.2ex}%
                                   {1.5ex \@plus .2ex}%
                                   {\normalfont\normalsize\it}}
\renewcommand\paragraph{\@startsection{paragraph}{4}{\z@}%
                                   {-3.25ex\@plus -1ex \@minus -.2ex}%
                                   {1.5ex \@plus .2ex}%
                                   {\normalfont\normalsize\bf}}

\numberwithin{equation}{section}

\def\revise#1       {\raisebox{-0em}{\rule{3pt}{1em}}%
                     \marginpar{\raisebox{.5em}{\vrule width3pt\
                     \vrule width0pt height 0pt depth0.5em
                     \hbox to 0cm{\hspace{0cm}{%
                     \parbox[t]{4em}{\raggedright\footnotesize{#1}}}\hss}}}}

\newcommand{\ie}{{\it i.e.,}\ }

\def\calc         {{\cal C}}

\def\cale         {{\cal E}}
\def\calf         {{\cal F}}

\def\calm         {{\cal M}}
\def\caln         {{\cal N}}
\def\calo         {{\cal O}}

\def\del          {\partial}

\def\sqr#1#2{{\vcenter{\vbox{\hrule height.#2pt
 \hbox{\vrule width.#2pt height#1pt \kern#1pt
 \vrule width.#2pt}\hrule height.#2pt}}}}

\newcommand{\kk}{\mathfrak{q}}
\newcommand{\ww}{\mathfrak{w}}

\def\a{\alpha}
\def\b{\beta}
\def\w{\omega}

\def\e{\epsilon}

\def\g{\gamma}

\def\aa1{\phi}
\def\cc1{\psi}

\def\arctanh{{\rm arctanh}}

\def\l{\lambda}

\def\hP{\hat{P}}

\def\hb{\hat{\beta}}

\catcode`\@=12

\begin{document}


\title{\bf  Violation of the holographic bulk viscosity bound}
\pubnum{UWO-TH-11/11}

\date{September 30, 2011}

\author{
Alex Buchel\\[0.4cm]
\it Department of Applied Mathematics\\
\it University of Western Ontario\\
\it London, Ontario N6A 5B7, Canada\\[0.2cm]
\it Perimeter Institute for Theoretical Physics\\
\it Waterloo, Ontario N2J 2W9, Canada\\[0.2cm]
 }

\Abstract{
Motivated by gauge theory/string theory correspondence, a lower bound on the
bulk viscosity of strongly coupled gauge theory plasma was proposed
in \cite{bb}. We consider strongly coupled $\caln=4$ supersymmetric
Yang-Mills plasma compactified on a two-manifold of constant curvature
$\b$.  We show that the effective $(1+1)$-dimensional hydrodynamic
description of the system is governed by the bulk viscosity violating
the bound of \cite{bb}, once $\b<0$.
 }

\makepapertitle

\body

\version\versionno
\tableofcontents

\section{Introduction}
Relativistic hydrodynamics is an effective description of a nearly-equilibrium system 
at time  (length) scales that are much longer (larger) than any characteristic 
microscopic scale of a system. In the simplest case (with no conserved charges), 
the dynamics of the hydrodynamic fluctuations in the system on a $(p+1)$-dimensional 
manifold $\calm_{p+1}$ with the metric 
\begin{equation}
ds^2=G_{\mu\nu} dx^\mu dx^\nu\,,
\eqlabel{metric}
\end{equation}
is governed by conservation of the 
stress-energy tensor $T_{\mu\nu}$,
\begin{equation}
\nabla_\nu T^{\mu\nu}=0\,.
\eqlabel{nt}
\end{equation}
The stress-energy tensor includes both and equilibrium part (with local energy density $\cale$ and pressure $P$) 
and a dissipative part $\Pi^{\mu\nu}$,
\begin{equation}
T^{\mu \nu} =\cale\, u^{\mu}u^{\nu}+P \Delta^{\mu \nu} +\Pi^{\mu
\nu}\quad {\rm where}\ \ \Delta^{\mu \nu}=G^{\mu
\nu}+u^{\mu}u^{\nu}\,.
\eqlabel{1.3}
\end{equation}
Above, $u^\mu$ is the local $(p+1)$-velocity of the fluid with $u^\mu
u_\mu=-1$. Further, $\Pi^{\mu\nu} u_\nu = 0$. In phenomenological
hydrodynamics, the dissipative term $\Pi^{\mu\nu}$,
\begin{equation}
\Pi^{\mu\nu}=\sum_{n=1}^\infty \Pi_n^{\mu\nu}\,,
\eqlabel{dissPi}
\end{equation}
 can be
represented as an infinite series expansion in velocity gradients
(and curvatures, for a fluid in a curved background), with the
coefficients of the expansion commonly referred to as transport
coefficients. In \eqref{dissPi} the subscript $\ _n$ denotes the total number 
of the velocity $\del_{\a_{1}}\cdots\del_{\a_i} u_\b$ and/or the background  metric $\del_{\a_1}\cdots\del_{\a_j} G_{\a\g}$ derivatives.
The familiar example of the Navier-Stokes equations
is obtained by truncating $\Pi^{\mu\nu}$ at linear order in this
expansion
\begin{equation}
\Pi^{\mu \nu}=\Pi_1^{\mu\nu}\left(\eta,\zeta\right)=-\eta\,
\sigma^{\mu \nu} -\zeta\, \Delta^{\mu \nu}\,\nabla\!\cdot\! u\,,
\eqlabel{1.4}
\end{equation}
where (for $p>1$)
\begin{equation}
\sigma^{\mu \nu}=2 \nabla^{\langle\mu}u^{\nu\rangle}\equiv
\Delta^{\mu \alpha} \Delta^{\nu \beta} \left(
\nabla_{\alpha}u_{\beta}+\nabla_{\beta}u_{\alpha}\right)
-\frac{2}{p}\Delta^{\mu \nu}\left(\Delta^{\alpha
\beta}\nabla_{\alpha}u_{\beta}\right)\,. \eqlabel{1.5}
\end{equation}
Notice that at this order in the hydrodynamic approximation we need
to introduce only two transport coefficients, namely the 
shear\footnote{To define the shear viscosity one needs $p>1$.}
$\eta$ and bulk $\zeta$ viscosities. 
At the second order in the derivative expansion,  $\Pi_2^{\mu\nu}$, one needs to introduce 
five \cite{ss1} or thirteen \cite{rom} additional transport coefficients (depending on whether or not the system is conformal).
 
Holographic gauge theory/string theory correspondence \cite{juan} provides a useful general guidance about 
hydrodynamic transport coefficients. Thus, in \cite{u2}, partly motivated by this correspondence, 
the authors (KSS) proposed a bound on the ratio of the shear viscosity 
to the entropy ratio\footnote{We use $\hbar=k_B=1$.} 
\begin{equation}
\frac{\eta}{s}\ge \frac{1}{4\pi}\,.
\eqlabel{shearb}
\end{equation}
The KSS bound is obeyed in all known fluids in Nature. It is either saturated or satisfied in all 
explicit or phenomenological examples of  holographic  gauge theory plasma at infinite coupling 
\cite{u2,u1,u5,u3,u4}. While the bound survives the leading order 't Hooft coupling corrections in four-dimensional 
holographic conformal models 
\cite{f1,f2,f3}, it can be violated in strongly coupled conformal gauge theories with fundamental matter\footnote{More generally,
in four-dimensional conformal theories with different central charges: $c\ne a$.}
\cite{vio1,vio2,vio3}\footnote{See \cite{vio4,bf1,bf2,bf3,bf4} for further discussion of the shear viscosity bound.}.

In \cite{bb} we proposed a bound on the bulk viscosity\footnote{See \cite{rel1,rel2,rel3} 
for the bound (and it violation) of a  particular 
second-order transport coefficient --- ``the effective relaxation time''. } 
of (infinitely) strongly coupled gauge theory plasma
\begin{equation}
\frac{\zeta}{\eta}\ge 2\left(\frac 1p -c_s^2\right)\,,
\eqlabel{bulk1}
\end{equation}
or equivalently (using the universality of the holographic shear viscosity),
\begin{equation}
\frac{\zeta}{s}\ge \frac{1}{2\pi}\left(\frac 1p -c_s^2\right)\,,
\eqlabel{bulk2}
\end{equation}
where $c_s^2$ is the speed of sound waves in plasma. In what follows we use the second version of the 
bound, eq.~\eqref{bulk2}, as it is applicable even for $p=1$. The bulk viscosity bound 
\eqref{bulk2} is saturated in toroidal compactifications of conformal theories \cite{bb,sk1}.
It is satisfied in various examples of string theory embedding of the holographic gauge theory/gravity 
correspondence: the strongly coupled $\caln=2^*$ gauge theory plasma \cite{n21,n22}, the cascading gauge theory 
plasma \cite{ca1,ca2}, the mass-deformed $\caln=4$ SYM plasma with a non-vanishing $U(1)$ R-charge chemical potential
\cite{mn4}. The bulk viscosity bound \eqref{bulk2} is further satisfied in some phenomenological models 
of the holographic correspondence \cite{ph1,ph2}, however, it is violated in some other phenomenological model 
\cite{ph4,ph3}.

In this paper we address the question whether the violation of the bulk viscosity bound \eqref{bulk2} is limited to the 
phenomenological models of the holographic gauge/gravity correspondence. A natural first guess is to replicate the 
strategy in the study of the KSS shear viscosity bound \eqref{shearb}, \ie to consider the effect of the gauge theory 
finite coupling and non-planar corrections on the purported bound.  Unfortunately, such an approach can not lead 
to a reliable conclusion. Indeed, when the bulk viscosity originates from toroidal compactifications of conformal hydrodynamics 
(in which case the string theory dual corrections to  finite coupling/non-planar corrections  can be under control),
the general arguments in \cite{bb} guarantee what the bulk viscosity bound continues to be saturated. 
On the other hand, a five-dimensional gravitational description of a generic non-conformal holographic correspondence involves 
scalar fields that originate from 3-form RR fluxes of type IIB supergravity --- in this case the 
full set of the higher derivative corrections to the supergravity is unknown. Clearly, we need to focus on 
explicit string theory examples of the holographic correspondence in the supergravity approximation.  

Our strategy in exploring the bound \eqref{bulk2} is to consider compactifications of the strongly coupled conformal
hydrodynamics on $k$-dimensional spatial manifolds $\calm_{k}$, with $k\le(p-1)$, of constant curvature $\b\ne 0$. 
Specifically, we consider hydrodynamic transport of strongly coupled $\caln=4$ $SU(N)$ SYM plasma on $\calm_2$,
\begin{equation}
\calm_2=\begin{cases}
S^2\,,\qquad &\Longrightarrow\qquad \b=+\frac{2}{L^2}\\
\Sigma^2=H^2/G\,, \qquad  &\Longrightarrow\qquad \b=-\frac{2}{L^2}
\end{cases}\,,
\eqlabel{defm2}
\end{equation}
where $S^2$ is a round two-sphere, and $H^2$ is  a hyperbolic space and $G$ is a discrete subgroup of its $SL(2,R)$ symmetry group.
The  quotient $\Sigma^2$ is assumed to be smooth and compact.      
$L$ is a "radius'' of $\calm_2$. Since we are left with a single infinitely large spatial direction, the low-energy effective 
description of the compactification is given by $(1+1)$-dimensional hydrodynamics. Notice that the latter hydrodynamic description 
{\it can not} be obtained from the compactification of the curved-space $(3+1)$-dimensional hydrodynamics of $\caln=4$ plasma ---
the gradients of the background metric along the $\calm_2$ directions are generically\footnote{A characteristic dimensionless 
parameter here is $\frac{\b}{T^2}$, while a typical hydrodynamic parameter is $\frac{\w}{T}$ or $\frac{|\vec{q}|}{T}$.
} large in the hydrodynamic 
limit, which 
invalidates the gradient expansion \eqref{dissPi}. Running ahead, we find that to leading order in $\frac{\b}{T^2}\ll 1$, 
the $\calm_2$-compactified hydrodynamics is affected by the {\it third}-order $\caln=4$ SYM hydrodynamics on 
$\calm_4=R^{1,1}\times \calm_2$.  To first-order in the velocity gradients, the effective $(1+1)$-dimensional hydrodynamic 
description of the theory is determined by a single transport coefficient --- the bulk viscosity $\zeta$. 
We extract $\zeta$ from the dispersion relation of the sound waves in $\caln=4$ plasma on $R^{1,1}\times \calm_2$ 
propagating along the single uncompactified spatial direction, which we denote as '$z$',
\begin{equation}
\ww=c_s^2\ \kk-i\ \pi \frac{\zeta}{s}\ \kk^2+\calo(\kk^3)\,,
\eqlabel{sound11}
\end{equation}
where 
\begin{equation}
\ww=\frac{\w}{2\pi T}\,,\qquad \kk=\frac{q_z}{2\pi T}\,,
\eqlabel{wk}
\end{equation}
and $s$ is the entropy density. The speed of sound waves is determined from the equation of state as follows
\begin{equation}
c_s^2=\frac{\del P_{zz}}{\del \cale}\bigg|_{\b={\rm const}}\,,
\eqlabel{cs2}
\end{equation}
where $\cale$ is the energy density, and $P_{zz}\equiv T_{zz}$ is the equilibrium pressure in the plasma 
in the $z$-direction.

The rest of the paper is organized as follows. In section \ref{thermo} we  discuss the regular black hole solution 
in $AdS_5$ with the asymptotic boundary metric $R^{1,1}\times \calm_2$, dual to an equilibrium thermal 
state of $\caln=4$ SYM plasma compactified on $\calm_2$. We compute the one-point correlation function of the 
boundary stress-energy tensor and  find the general expression for the speed of the sound waves \eqref{cs2}. 
For general values of $\frac{\b}{T^2}$, the background geometry (and its thermodynamics) is found numerically. 
We present analytical results for the thermodynamics to leading order in  $\frac{\b}{T^2}\ll 1$.
In section \ref{hydro} we compute the dispersion relation of the sound channel quasinormal modes  
\cite{ks} in the black hole geometry of section \ref{thermo}, and extract the speed of the sound waves 
$c_s^2$ as well as the sound waves attenuation coefficient $\Gamma=\pi \frac{\zeta}{s}$. 
We present analytic results to leading order at high temperatures, $\frac{\b}{T^2}\ll 1$, and 
numerical results for generic values of  $\frac{\b}{T^2}$. In section \ref{resbound} we discuss the 
bulk viscosity bound and its violation in $\caln=4$ SYM plasma on $R^{1,1}\times \calm_2$. 
We conclude in section \ref{conclude}.

\section{Thermodynamics of $\caln=4$ SYM plasma on $R^{1,1}\times \calm_2$}\label{thermo}
Effective five-dimensional gravitational action describing $\caln=4$ $SU(N)$ SYM on $\calm_4=R^{1,1}\times \calm_2$
in the planar limit  ($g_{YM}\to 0$, $N\to \infty$ with $\l\equiv g_{YM}^2N={\rm const}$), and for an infinite 
't Hooft coupling $\l\to \infty$, is given by  
\begin{equation}
S_5=\frac{1}{16\pi G_5}\int_{\calm_5,\ \del\calm_5=\calm_4}\ d^5\xi\sqrt{-g}\left(R+12\right)\,,
\eqlabel{5dac}
\end{equation}
where without loss of generality we normalized the radius of curvature of asymptotically $AdS_5$ geometry to $1$. 
Such a normalization implies 
\begin{equation}
G_5=\frac{\pi}{2 N^2}\,.
\eqlabel{g5qft}
\end{equation}

The background geometry dual to a state of the theory with translational invariance along the $z$-direction and 
the symmetry of $\calm_2$ is given by
\begin{equation}
ds_5^2=-c_1^2\ dt^2+c_2^2\ \frac{2}{\b}\ (d\calm_2)^2+c_3^2\ dz^2+c_4^2\ dr^2\,,
\eqlabel{5dmet}
\end{equation}
where $c_i=c_i(r)$ and $ (d\calm_2)^2$ is the standard metric on $\calm_2$.
Given \eqref{5dmet} we find the following second order equations of motion
\begin{equation}
\begin{split}
0=&c_1''-\frac{c_1(c_2')^2}{3c_2^2} +c_1' \left(
\frac{4c_2'}{3c_2} -\frac{c_4'}{c_4}+\frac{2c_3'}{3c_3}\right)
- \frac{2c_1c_3'c_2'}{3c_2c_3}-2 c_1 c_4^2+\frac\b6 \frac{c_4^2 c_1 }{c_2^2}\,,
\end{split}
\eqlabel{bac1}
\end{equation}
\begin{equation}
\begin{split}
0=&c_2''+\frac{2 (c_2')^2}{3 c_2}+c_1' \left(
\frac{c_2'}{3c_1}-\frac{c_3' c_2}{3c_3 c_1}\right)
+c_2' \left(  \frac{c_3'}{3c_3}-\frac{c_4'}{c_4}\right)-2 c_4^2 c_2-\frac\b3 \frac{c_4^2} {c_2}\,,
\end{split}
\eqlabel{bac2}
\end{equation}
\begin{equation}
\begin{split}
0=&c_3''-\frac{(c_2')^2c_3}{3c_2^2} -\frac 23 c_1' \left( \frac{c_2' c_3}{c_2 c_1}- \frac{c_3'}{c_1}\right)
+\left(\frac{4c_2'}{3c_2}-\frac{c_4'}{c_4}\right) c_3'-2 c_4^2 c_3+\frac\b6 \frac{c_3 c_4^2} {c_2^2}\,,
\end{split}
\eqlabel{bac3}
\end{equation}
as well as the first order constraint
\begin{equation}
\begin{split}
0=&2 \frac{c_2' c_1' c_2}{c_1}+(c_2')^2+\frac{c_1' c_3' c_2^2}{c_3 c_1}-\frac\b2 c_4^2 
-6 c_4^2 c_2^2+2 \frac{c_3' c_2' c_2}{c_3}\,.
\end{split}
\eqlabel{bac4}
\end{equation}
We explicitly verified that \eqref{bac4} is consistent with \eqref{bac1}-\eqref{bac3}.

\subsection{Asymptotics of the background geometry dual to a thermal state of the theory}

To describe a thermal state of strongly coupled $\caln=4$ SYM plasma on $R^{1,1}\times \calm_2$ we find 
it convenient to use the radial coordinate 
\begin{equation}
x\equiv 1-\frac{c_1}{c_3}\,,
\eqlabel{defx}
\end{equation}
and further introduce\footnote{The two second-order equations of motion for $\{a,g\}$ can be obtained 
from \eqref{bac1}-\eqref{bac4}. As they are not very illuminated, we do not present them here.}  
\begin{equation}
c_2(x)=\frac{a(x) g(x)}{(2x-x^2)^{1/4}}\,,\qquad c_3(x)=\frac{a(x)}{(2x-x^2)^{1/4}}\,.
\eqlabel{defga}
\end{equation}
Near the boundary, \ie as $x\to 0_+$, the asymptotics of $\{a,g\}$ are given by
\begin{equation}
\begin{split}
a=&\mu\left(1+\frac{\b\sqrt{2}}{18\mu^2}\ x^{1/2}+\left(a_{2,0}+\frac{\b^2}{324\mu^4}
-\frac{\b^2}{384\mu^4}\ \ln x\right)\ x+\calo\left(x^{3/2}\right)
\right)\,,\\
g=&1-\frac{\b\sqrt{2}}{8\mu^2}\ x^{1/2}+\left(\frac{127\b^2}{10368\mu^4}-2a_{2,0}+\frac{\b^2}{192\mu^4}\ \ln x\right)\ x
+\calo\left(x^{3/2}\right)\,.
\end{split}
\eqlabel{uvass}
\end{equation}
They are characterized by two independent parameters: $\{\mu\,, a_{2,0}\}$. 
As we will see later, $\mu$ is related to the temperature of the thermal state, and $a_{2,0}$ determines the 
one-point correlation function of the boundary stress-energy tensor at thermal equilibrium.
Near the regular horizon, \ie as $y\equiv (1-x)\to 0_+$, the asymptotics of $\{a\,, g\}$ are given by
\begin{equation}
\begin{split}
a=&\mu\biggl(a_0^h+\frac14{a_0^h\left(32g_1^hg_0^h {(a_0^h)^2}-1\right)}\ y^2+\calo(y^4)\biggr)\,,\\
g=&g_0^h+\frac{\b}{\mu^2} g_1^h\ y^2+\calo(y^4)\,.
\end{split}
\eqlabel{irass}
\end{equation}
They are characterized by three independent parameters: $\{a_0^h\,, g_0^h\,, g_1^h\}$.
Notice that, given $\mu$, there are precisely four parameters
\[
\{a_{2,0}\,, a_0^h\,, g_0^h\,, g_1^h\}\,,
\]
which is necessary to uniquely specify a solution for the second-order ODEs for $\{a,g\}$.

Given \eqref{irass} we can compute the Hawking temperature $T$ and the Bekenstein-Hawking 
entropy density $s$ of the black hole geometry \eqref{5dmet}
\begin{equation}
\left(\frac{2\pi T}{\mu}\right)^2=\frac{1}{8g_0^h g_1^h}\,,\qquad \frac{s}{\mu^3}=\frac{(a_0^h)^3g_0^h}{4 G_5}\,.
\eqlabel{gentem}
\end{equation}

For general $\frac{\b}{\mu^2}$, the equations of motion for $\{a\,, g\}$ are too complicated to be solved analytically ---
we solve them numerically.
Analytical solution is possible though to leading order in the dimensionless parameter $\frac{\b}{\mu^2}$.
Using the high-temperature parametrization,
\begin{equation}
a(x)=\mu\left(1+\frac{\b}{\mu^2}\ a_1(x)+\calo\left(\frac{\b^2}{\mu^4}\right)\right)\,,\qquad 
g(x)=1+\frac{\b}{\mu^2}\ g_1(x)+\calo\left(\frac{\b^2}{\mu^4}\right)\,,
\eqlabel{parhight}
\end{equation}
we find
\begin{equation}
\begin{split}
a_1(x)=&\frac{1+2x -x^2}{24 (2x-x^2)} \left(\ln(1-x)+\arctanh\sqrt{2x-x^2}\right)
+\frac{2x-x^2-2 \sqrt{2x-x^2}}{48 (2x-x^2)}\,,\\
g_1(x)=&-\frac18 \arctanh\sqrt{2x-x^2}-\frac 18 \ln(1-x)\,.
\end{split}
\eqlabel{leadbac}
\end{equation}
Given \eqref{leadbac}, we find\footnote{We get $\calo(\b/\mu^2)$ coefficient of $g_1^h$ we actually need to compute 
the second order correction, $g_2(x)$ in \eqref{parhight}. This is a straightforward extension and we omit the details.}
\begin{equation}
\begin{split}
a_{2,0}=&-\frac{1}{16}\ \frac{\b}{\mu^2}+\calo\left(\frac{\b^2}{\mu^4}\right)\,,\qquad a_0^h=\mu\left(1
+\frac{\b}{\mu^2}\ \left(\frac{1}{12}\ln2-\frac{1}{48}\right)+\calo\left(\frac{\b^2}{\mu^4}\right)\right)\,,\\
g_0^h=&1-\frac 18\ln 2\  \frac{\b}{\mu^2}+\calo\left(\frac{\b^2}{\mu^4}\right)\,,\qquad 
g_1^h=\frac{1}{32}+\frac{\ln 2-1}{256}\ \frac{\b}{\mu^2}+\calo\left(\frac{\b^2}{\mu^4}\right)\,.
\end{split}
\eqlabel{highTpar}
\end{equation}
which implies (see \eqref{gentem}) 
\begin{equation}
\frac{\pi T}{\mu}=1+\frac{1}{16}\ \frac{\b}{\mu^2}+\calo\left(\frac{\b^2}{\mu^4}\right)\,,\qquad
\frac{s}{\mu^3}=\frac{1}{4G_5}\ \left(1-\frac{1}{16}\ \frac{\b}{\mu^2}+\calo\left(\frac{\b^2}{\mu^4}\right)\right)\,.
\eqlabel{highttdef}
\end{equation}

\subsection{Holographic renormalization and the boundary stress-energy tensor}
Holographic renormalization of strongly coupled $\caln=4$ SYM plasma has been extensively discussed in the literature, 
see \cite{ps} for example. To render (dual gravitational) correlation functions finite, the effective action 
\eqref{5dac} has to be supplemented with the following 
set of counterterms:
\begin{equation}
S_{ct}=-\frac{3}{8\pi G_5}\int_{\calm_4, \e=c_3^{-2}}\ \sqrt{-\gamma}
\left(1+\frac 12 \hat{P} -\frac{1}{12}\left(\hP^{kl}\hP_{kl}-\hP^2\right)\ln\e\right)\,,
\eqlabel{sct}
\end{equation} 
where $\g$ is the metric \eqref{5dmet} restricted to $c_3^{-2}=\e$, and 
\begin{equation}
\hP=\g^{ij}\hP_{ij}\,,\qquad \hP_{ij}=\frac 12\left(R_{ij}-\frac 16 R \g_{ij}\right)\,.
\eqlabel{hpdef}
\end{equation}
In practice, we use expressions for the regularized one-point correlation function of the stress-energy tensor 
obtained in \cite{aby}, in particular, see eq.~(3.52), in the conformal limit. We find:
\begin{equation}
\begin{split}
8\pi G_5\ T_{tt}=&\frac{4\pi^2}{N^2}\ \cale=
\frac32\ \mu^4-2 \mu^4\ a_{2,0}+\frac{1}{48}\ \b^2\ \ln\frac{\mu}{\Lambda}\,,\\
8\pi G_5\ T_{zz}=&\frac{4\pi^2}{N^2}\ P_{zz}=\frac12\ \mu^4+2 \mu^4\ a_{2,0}-\frac{1}{48}\ \b^2\ 
\ln\frac{\mu}{\Lambda}\,,
\end{split}
\eqlabel{tij}
\end{equation}
where $\Lambda$ is an arbitrary (fixed) renormalization scale associated with the 
ambiguity\footnote{The same ambiguity  appears in computation of the thermal stress-energy tensor
of strongly coupled $\caln=2^*$  plasma \cite{bn2}.} of defining 
the stress-energy tensor of a theory on curved background manifold $\calm_4$ with broken supersymmetry.

Given \eqref{tij}, it is straightforward to compute the speed of the sound waves \eqref{cs2} in $\calm_2$-compactified 
hydrodynamics of $\caln=4$ SYM plasma:
\begin{equation}
c_s^2=\frac{{\del P_{zz}}}{{\del\cale}}\bigg|_{\b={\rm const}}
=\frac{1+4 a_{2,0}(\hb)-2\hb\ a_{2,0}'(\hb)-\frac{1}{96} \hb^2}{3-4 a_{2,0}(\hb)+2\hb\  a_{2,0}'(\hb) +\frac{1}{96} \hb^2}\,,
\qquad \hb\equiv\frac{\b}{\mu^2}\,,
\eqlabel{cs22th}
\end{equation}
where the prime denotes derivative with respect to $\hb$.
In section \ref{hydro} we compare the thermodynamic prediction \eqref{cs22th} for the speed of the sound waves with 
direct computations from the dispersion relation of the sound channel quasinormal modes.

Using the asymptotic expansions \eqref{highTpar} and \eqref{highttdef} we find
\begin{equation}
\begin{split}
\cale=&\frac{3\pi^2N^2 T^4}{8}\left(1-\frac{\b}{6 \pi^2 T^2}+\calo\left(\frac{\b^2}{T^4}\right)\right)\,,\\
P_{zz}=&\frac{\pi^2N^2 T^4}{8}\left(1-\frac{\b}{2 \pi^2 T^2}+\calo\left(\frac{\b^2}{T^4}\right)\right)\,,
\end{split}
\eqlabel{hightthermo1}
\end{equation}
and\footnote{To compute $c_s^2$ to order $\calo(\b^3/\mu^6)$ we do not need the $\calo(\b^2/\mu^4)$
coefficient in the expansion of $a_{2,0}$, see \eqref{highTpar}.}
\begin{equation}
c_s^2=\frac 13-\frac{\beta}{18\mu^2}-\frac{\beta^2}{432\mu^4}+\calo\left(\frac{\b^3}{\mu^6}\right)\,.
\eqlabel{hightthermo2}
\end{equation}
Notice from \eqref{highttdef} that since 
\begin{equation}
s=\frac{\pi^2N^2 T^3}{2}\left(1-\frac{\b}{4 \pi^2 T^2}+\calo\left(\frac{\b^2}{T^4}\right)\right)\,,
\eqlabel{shight}
\end{equation}
the basic thermodynamic relations 
\begin{equation}
-P_{zz}=\calf=\cale-T s\,,\qquad d\cale=T ds\,,
\eqlabel{thermorel}
\end{equation}
are (analytically) satisfied to order $\calo\left(\frac{\b^2}{T^4}\right)$.

\section{Sound channel quasinormal mode of $AdS_5$ black hole with $R^{1,1}\times \calm_2$ 
asymptotic boundary}\label{hydro}

Following \cite{ks}, the dispersion relation for the sound waves in $\calm_2$-compactified 
$\caln=4$ SYM plasma is identified with the dispersion relation for the sound channel quasinormal modes
propagating in the $z$-direction of the black hole geometry \eqref{5dmet}. We briefly outline the construction 
of the corresponding fluctuations. 

Consider the following decoupled set of metric fluctuations 
\begin{equation}
\begin{split}
g_{\mu\nu}\to& g_{\mu\nu}+h_{\mu\nu}\,,\\
h_{tt}(t,z,r)=&c_1^2(r)\ H_{tt}(r)\ e^{-i \w t+i q_z z}\,,\\
h_{tz}(t,z,r)=&c_3^2(r)\ H_{tz}(r)\ e^{-i \w t+i q_z z}\,,\\
h_{ij}(t,z,r)=&c_3^2(r)\ H_{ss}(r)\ e^{-i \w t+i q_z z}\ \g^{(\calm_2)}_{ij}\,,\\
h_{zz}(t,z,r)=&c_3^2(r)\ H_{zz}(r)\ e^{-i \w t+i q_z z}
\end{split}
\eqlabel{metricfluc}
\end{equation}
where 
\begin{equation}
\g^{(\calm_2)}_{ij}d\xi^id\xi^j
=\frac 2\b (d\calm_2)^2\,.
\eqlabel{defcalm2}
\end{equation}
From the equations of motion for the fluctuations \eqref{metricfluc} we obtain 4 second-order 
linear ODEs for $\{H_{tt},H_{tz},H_{zz},H_{ss}\}$ as well as 3 linear first-order constraints associated 
with the diffeomorphism-fixing conditions 
\begin{equation}
h_{tr}=h_{zr}=h_{rr}=0\,.
\eqlabel{diffeofix}
\end{equation}
The combination 
\begin{equation}
Z_H\equiv \frac{4 \kk}{\ww}\ H_{tz}+2 \ H_{zz}-2 \ H_{ss} \
\left(\frac{c_3 c_3'}{c_2 c_2'}-\frac{\kk^2}{\ww^2} \frac{c_1 c_1'}{c_2c_2'}\right)
+2 \frac{\kk^2}{\ww^2}\ {c_1^2}{c_3^2}\ H_{tt} \,,
\eqlabel{defz}
\end{equation}
is invariant under the residual diffeomorphisms (for the gauge fixing \eqref{diffeofix}) and 
satisfies the following decoupled linear ODE\footnote{The expressions for $\calc_i$ 
are too long to be presented here --- they are available from the author upon request. }
\begin{equation}
0=Z_H''+\calc_1\ Z_H'+\calc_2\ Z_H\,,\qquad \calc_i=\calc_i\biggl[\{c_{1,2,3}(r)\},\ww,\kk\biggr]\,.
\eqlabel{zheq}
\end{equation} 
In terms of the radial coordinate \eqref{defx}, the hydrodynamic limit takes form
\begin{equation}
Z_H(x)=(1-x)^{-i\ww}\biggl(z_{H,0}(x)+i\kk\ z_{H,1}(x)+\calo(\kk^2)\biggr)\,,\qquad \ww=c_s\ \kk-i\Gamma\ \kk^2+\calo(\kk^3)\,,
\eqlabel{hydlimit}
\end{equation}
with the following boundary conditions
\begin{equation}
\begin{split}
&\lim_{x\to 1_-}z_{H,0}=1\,,\qquad \lim_{x\to 1_-}z_{H,1}=0\,,\\
&z_{H,i}=\calo(x)\,,\qquad {\rm as}\qquad x\to 0_+\,.
\end{split}
\eqlabel{bc}
\end{equation}

We omit the details of solving the boundary value problem \eqref{zheq}-\eqref{bc} 
for $\{z_{H,0}, z_{H,1}\}$ and present only the 
results\footnote{For more details on the solution procedure see \cite{n22}.}. 
First of all, we  find 
\begin{equation}
\begin{split}
\pm\ c_s=&\frac {1}{\sqrt{3}}-\frac{\sqrt{3}}{36}\ \frac{\b}{\mu^2}-0.004009(4)\ \frac{\b^2}{\mu^4}
+\calo\left(\frac{\b^3}{\mu^6}\right)\,,\\
\Gamma=&\frac 13+\frac{\ln2}{18}\ \frac{\b}{\mu^2}+\calo\left(\frac{\b^2}{\mu^4}\right)\,.
\end{split}
\eqlabel{highthydro}
\end{equation}
Notice that the speed of the sound waves is in perfect agreement with the predictions from the thermodynamics
\eqref{hightthermo2} --- the coefficient of $\frac{\b^2}{\mu^4}$ agrees with an accuracy of $\propto 10^{-11}$.

\begin{figure}[t]
\begin{center}
\psfrag{cs2}{{$c_s^2$}}
\psfrag{b2piT2}{{$\frac{\b}{(2\pi T)^2}$}}
\includegraphics[width=2.8in]{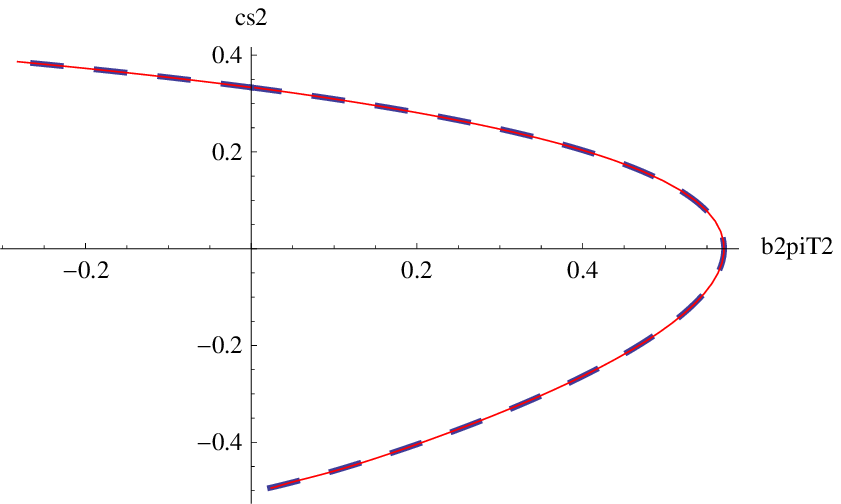}
\includegraphics[width=2.8in]{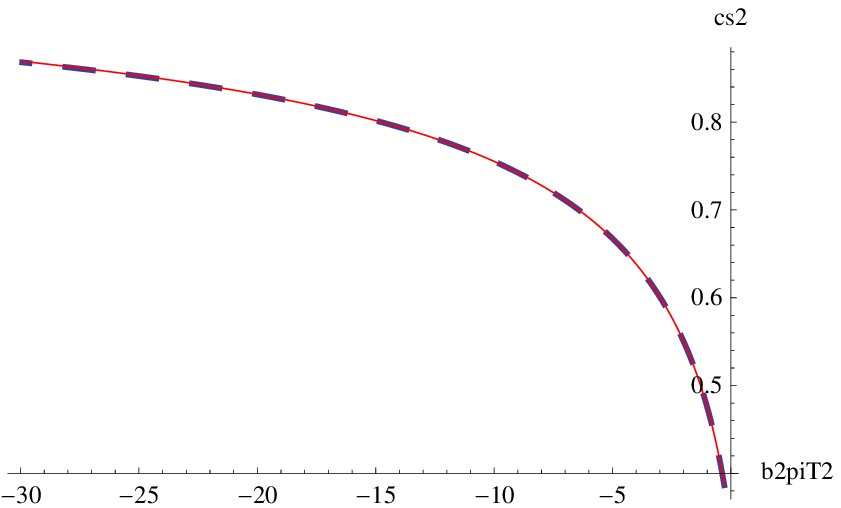}
\end{center}
  \caption{(Colour online) The speed of the sound waves in $\calm_2$-compactified $\caln=4$ SYM plasma 
as a function of $\frac{\b}{(2\pi T)^2}$. The dashed blue curves indicate the speed of the sound obtained 
from the quasinormal mode dispersion relation, see \eqref{hydlimit}. The solid red curves 
indicate the prediction for the speed of the sound waves from the equilibrium thermodynamics, 
see \eqref{cs22th}. 
} \label{figure1}
\end{figure}

\begin{figure}[t]
\begin{center}
\psfrag{cs2}{{$c_s^2$}}
\psfrag{G}{{$\Gamma$}}
\psfrag{b2piT2}[b][b]{{$\frac{\b}{(2\pi T)^2}$}}
\includegraphics[width=2.8in]{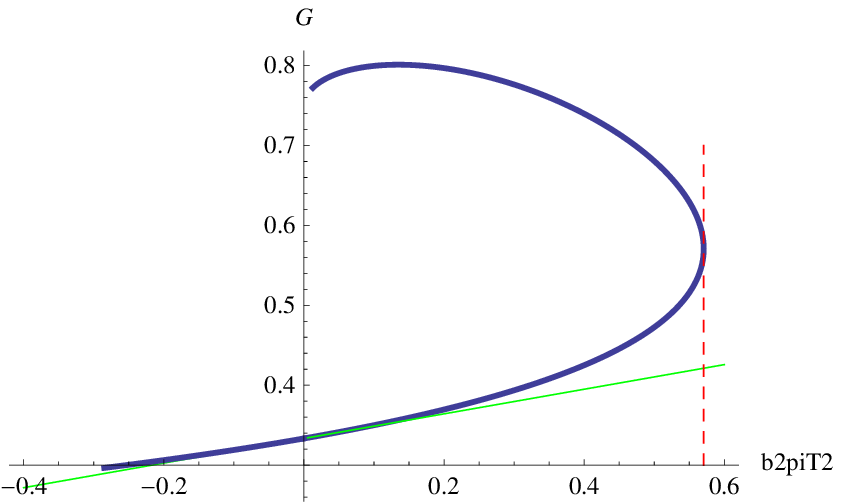}
\includegraphics[width=2.8in]{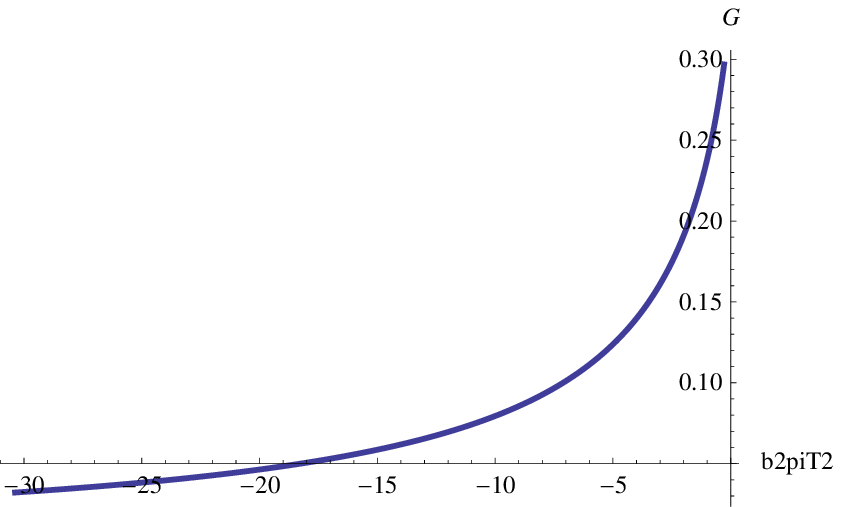}
\end{center}
  \caption{(Colour online) The attenuation  of the sound waves (solid blue lines), $\Gamma=\pi\frac{\zeta}{s}$, 
in $\calm_2$-compactified $\caln=4$ SYM plasma 
as a function of $\frac{\b}{(2\pi T)^2}$.  The solid green line 
indicates the high-temperature prediction  
\eqref{highthydro}. The vertical dashed red line indicates the critical temperature $T_c$ at which 
the speed of the sound waves vanishes.
} \label{figure2}
\end{figure}

Figure~\ref{figure1} presents the speed of the sound waves in $\calm_2$-compactified $\caln=4$ SYM
plasma as a function of  $\frac{\b}{(2\pi T)^2}$. The dashed blue curves are obtained from the dispersion
relation of the sound quasinormal modes \eqref{hydlimit}, while the solid red curves indicate the 
prediction    for the speed of the sound waves from the equilibrium thermodynamics, 
see \eqref{cs22th}. Notice that at 
\begin{equation}
\frac{\b}{(2\pi T_c)^2}=0.570580(8)\,,\qquad c_s^2\bigg|_{T=T_c}=0\,,
\eqlabel{ttcdef}
\end{equation}
the speed of the sound waves squared vanishes, and continuous to the unstable branch with $c_s^2<0$.
In the vicinity of the critical point $c_s^2\propto \left(1-\frac{T_c}{T}\right)^{1/2}$ --- the same critical
phenomena has been observed in $\caln=2^*$ gauge theory plasma \cite{bn2}, and the cascading gauge theory 
plasma \cite{ca2}.  

Figure~\ref{figure2} presents the attenuation $\Gamma=\pi\frac{\xi}{s}$
of the sound waves in $\calm_2$-compactified $\caln=4$ SYM
plasma as a function of  $\frac{\b}{(2\pi T)^2}$ (solid blue curves).  The solid green line 
indicates the high-temperature prediction  
\eqref{highthydro}. The vertical dashed red line indicates the critical temperature $T_c$, 
see \eqref{ttcdef}. Notice that as in the case of $\caln=2^*$ gauge theory plasma \cite{bb}, the bulk viscosity remains 
finite at the critical point.

\section{Bulk viscosity bound and its violation in $\caln=4$ SYM plasma on $R^{1,1}\times \calm_2$}\label{resbound}

\begin{figure}[t]
\begin{center}
\psfrag{cs2}{{$\left(\frac 13-c_s^2\right)$}}
\psfrag{l}{{$\lambda$}}
\includegraphics[width=4in]{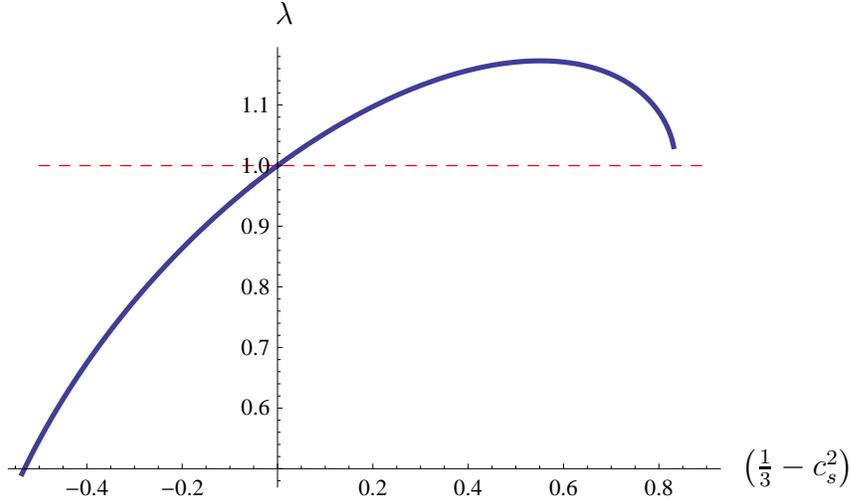}
\end{center}
  \caption{(Colour online) Violation of the bulk viscosity bound $\l\ge 1$ (see \eqref{boundfin}) in 
$\caln=4$ SYM plasma compactified on $\calm_2$ as a function of the 
conformal symmetry breaking parameter $\left(\frac 13-c_s^2\right)$.
} \label{figure3}
\end{figure}

The bulk viscosity bound proposed in \cite{bb}, in the case of strongly coupled $\caln=4$ SYM plasma 
compactified on $\calm_2$ (see \eqref{defm2})
reads 
\begin{equation}
\frac{\xi}{s}= \frac {\l}{2\pi} \left(1-c_s^2\right)\,,\qquad \l\ge 1 \,.
\eqlabel{boundfin}
\end{equation}
From \eqref{highthydro} we find
\begin{equation}
\begin{split}
\l=&1+\frac{2\ln 2-1}{3}\ \frac{\b}{(2\pi T)^2}+\calo\left(\frac{\b^2}{(2\pi T)^4}\right)\,.
\end{split}
\eqlabel{lpert}
\end{equation}
Clearly, if $\b<0$, \ie strongly coupled $\caln=4$ SYM plasma is compactified on $\Sigma^2$,
the bulk viscosity bound is violated. The leading (at high-temperature or equivalently small 
$\calm_2$ curvature  compactification) violation of the bulk viscosity bound is $\propto \b\propto R$. 
It comes from 2-derivatives of the metric along $\calm_2$ directions.  When viewed from the 
perspective of higher-order hydrodynamics of $\caln=4$ SYM plasma on $\calm_4=R^{1,1}\times \calm_2$,
such a violation comes from the {\it third}-order dissipative term, $\Pi^{\mu\nu}_3$.  

Figure~\ref{figure3} presents parameter $\l$ in the bulk viscosity bound \eqref{boundfin}
as a function of the conformal symmetry breaking parameter $\left(\frac 13-c_s^2\right)$ 
of strongly coupled $\caln=4$ SYM plasma compactified on $\calm_2$. Notice that $\l<1$ always, 
as long as $\b<0$, \ie the compactification manifold $\calm_2$ is a constant  curvature 
 higher genus Riemann surface. The violation is rather strong at low temperatures: it is 
$\approx 51\%$ at the lowest temperature we accessed numerically 
\begin{equation}
\min \left[\frac{\b}{(2\pi T)^2}\right]=-30.379\,.
\eqlabel{lowtem}
\end{equation}

\section{Conclusion}\label{conclude}

In this paper we constructed a specific string-theoretic counter-example to the bulk viscosity 
bound in (infinitely) strongly coupled gauge theory plasma proposed in \cite{bb}. 
We observed that compactification of the higher-dimensional  hydrodynamics on curved 
manifolds results in Navier-Stokes (first-order) hydrodynamics  with transport coefficients 
that are sensitive to higher-order dissipative terms of the higher-dimensional  hydrodynamics.
In the small-curvature limit of compactifications, the sensitivity starts with the 
third-order dissipative terms of the higher-dimensional conformal hydrodynamics.
Since flat-space compactifications of the conformal hydrodynamics saturate the bulk viscosity bound
of the effective lower-dimensional hydrodynamics \cite{bb}, conformal hydrodynamics compactifications 
on curved manifolds are guaranteed to  violate the bound
for the judicious choice of the compactification manifold curvature. 
In a specific example of strongly coupled $\caln=4$ SYM plasma we showed that violation occurs 
for compactifications on negative curvature two-manifolds. 

Since $\caln=4$ SYM contains conformally coupled scalars, one wonders whether the violation
of the bulk viscosity bound can be attributed to the presence of 
tachyons\footnote{Tachyons do not immediately imply the instabilities in the theory.} 
in the theory, when compactified 
on $\Sigma^2$. We do not believe this to be the case: in $\caln=2^*$ gauge theory plasma 
\cite{bn2} one can study mass deformations with $\frac{m_b^2}{T^2}<0$; such deformations do not violate the 
bound \cite{bb} since, for example at high-temperatures,  the bulk viscosity is affected 
at order  $\frac{m_b^4}{T^4}$. 

We did not explore in details the issue of stability of $\calm_2$-compactifications  of 
$\caln=4$ SYM discussed here. We did verify though that, at least at high temperature,
 there are no instabilities of metric 
fluctuations whose wave-functions are constant over $\calm_2$. Likewise, there are no instabilities 
of minimally coupled scalar (for example, a dilaton) in the background geometry \eqref{5dmet},
again, provided its wave-function is constant over $\calm_2$. In case of $\Sigma^2$ compactifications
(which lead to the violation of the bulk viscosity bound) fluctuations with non-trivial wave-function
on $H^2$ might be projected by the action of $G$ on the quotient $\Sigma^2=H^2/G$.
The latter fact prevents making a general statement about the stability of the $\Sigma^2$
compactifications.

Finally, in this paper we considered compactifications of $\caln=4$ SYM on $\calm_2$ which completely 
break the supersymmetry. It would be interesting to extend analysis presented here to supersymmetric 
(twisted) compactifications of  $\caln=4$ SYM  discussed in \cite{mmnn}.

\section*{Acknowledgments}
I would like to thank 
Rob Myers for valuable discussions.   Research at Perimeter
Institute is supported by the Government of Canada through Industry
Canada and by the Province of Ontario through the Ministry of
Research \& Innovation. I gratefully acknowledge further support by an
NSERC Discovery grant.

\end{document}